\documentclass[aps,pre,superscriptaddress,showpacs,twocolumn]{revtex4}

\usepackage{amsfonts}
\usepackage{amsmath}
\usepackage{multirow}
\usepackage{longtable}
\usepackage{graphicx}
\usepackage{color}

\newcommand{\be}{\begin{equation}}
\newcommand{\ee}{\end{equation}}
\newcommand{\bea}{\begin{eqnarray}}
\newcommand{\eea}{\end{eqnarray}}
\newcommand{\av}[1]{\langle #1 \rangle}
\newcommand{\bb}[1]{\mathbf{ #1}}
\begin{document}

\title{Epidemic spreading on time-varying multiplex networks}

\author{Quan-Hui Liu}
\affiliation{Web Science Center, University of Electronic Science and Technology of China, Chengdu 611731, China}
\affiliation{Big Data Research Center, University of Electronic Science and Technology of China, Chengdu 611731, China}
\affiliation{Laboratory for the Modelling of Biological and Socio-technical Systems, Northeastern University, Boston MA 02115 USA}

\author{Xinyue Xiong}
\affiliation{Laboratory for the Modelling of Biological and Socio-technical Systems, Northeastern University, Boston MA 02115 USA}

\author{Qian Zhang}
\affiliation{Laboratory for the Modelling of Biological and Socio-technical Systems, Northeastern University, Boston MA 02115 USA}

\author{Nicola Perra}
\email[]{n.perra@greenwich.ac.uk}
\affiliation{Centre for Business Network Analysis, University of Greenwich, Park Row, London SE10 9LS, United Kingdom}


\date{\today}

\begin{abstract}
Social interactions are stratified in multiple contexts and are subject to complex temporal dynamics.  The systematic study of these two features of social systems has started only very recently mainly thanks to the development of multiplex and time-varying networks. However, these two advancements have progressed almost in parallel with very little overlap. Thus, the interplay between multiplexity and the temporal nature of connectivity patterns is poorly understood. Here, we aim to tackle this limitation by introducing a time-varying model of multiplex networks. We are interested in characterizing how these two properties affect contagion processes. To this end, we study SIS epidemic models unfolding at comparable time-scale respect to the evolution of the multiplex network. We study both analytically and numerically the epidemic threshold as a function of the overlap between, and the features of, each layer. We found that, the overlap between layers significantly reduces the epidemic threshold especially when the temporal activation patterns of overlapping nodes are positively correlated. Furthermore, when the average connectivity across layers is very different, the contagion dynamics are driven by the features of the more densely connected layer. Here, the epidemic threshold is equivalent to that of a single layered graph and the impact of the disease, in the layer driving the contagion, is independent of the overlap. However, this is not the case in the other layers where the spreading dynamics are sharply influenced by it. The results presented provide another step towards the characterization of the properties of real networks and their effects on contagion phenomena
\end{abstract}

\maketitle

Social interactions take place in different contexts and modes of communication. On a daily basis we interact at work, in the family and across a wide range of online platforms or tools, e.g. Facebook, Twitter, emails, mobile phones etc. In the language of modern Network Science, social networks can be conveniently modeled and described as multilayers networks~\cite{ginestra18-1,kivela2014multilayer,de2013mathematical,boccaletti2014structure,cozzo2018multiplex}. This is not a new idea. Indeed, the concept of \emph{multiplexity} to describe the stratification of interactions dates back several decades \cite{wasserman1994social,verbrugge1979multiplexity,doi:10.1086/228943}. However, the digitalization of our communications and the miniaturization of devices has just recently provided the data necessary to observe, at scale, and characterize the multilayer nature of social interactions. \\
{\indent}As in the study of single layered networks, the research on multilayer graphs is divided in two interconnected areas. The first deals with the characterization of the structural properties of such entities~\cite{boccaletti2014structure,ginestra18-1}.  One of the central observations is that the complex topology describing each type of interactions (i.e., each layer) might be different. Indeed, the set and intensity of interactions in different contexts (e.g., work, family etc..) or platforms (e.g., Facebook, Twitter etc..) is not the same. Nevertheless, we are present in some, if not all, these layers which are then coupled by different degrees of overlap. Another interesting feature of multilayer graphs is that the connectivity patterns in different layers might be topologically and temporally correlated~\cite{starnini2017effects}. The second area of research instead considers the function, such as sustaining diffusion or contagion processes, of multilayer networks~\cite{ginestra18-1,de2016physics, salehi2015spreading}. A large fraction of this research aims at characterizing how the complex structural properties of multilayer graphs affect dynamical processes unfolding on their fabric. The first important observation is that disentangling connections in different layers, thus acknowledging multiplexity, gives raise to complex and highly non-trivial dynamics function of the interplay between inter and intra-layer connections~\cite{gomez2013diffusion,huang2018global,bianconi2017epidemic,osat2017optimal,jiang2018resource,granell2013dynamical,massaro2014epidemic,zhao2014multiple,buono2014epidemics,zhao2014immunization,wang2014asymmetrically,hu2014conditions,cozzo2013contact}.  A complete summary of the main results in the literature is beyond the scope of the paper. We refer the interested reader to these recent resources for details~\cite{ginestra18-1,kivela2014multilayer,de2013mathematical,boccaletti2014structure,cozzo2018multiplex}. \\
{\indent}Despite the incredible growth of this area of Network Science over the last years, one particular aspect of multilayer networks is still largely unexplored: the interplay between multiplexity and the temporal nature of the connectivity patterns especially when dynamical processes unfolding on their fabric are concerned~\cite{salehi2015spreading}. This should not come as a surprise. Indeed, the systematic study of the temporal dynamics even in single layered graphs is very recent. In fact, the literature has been mostly focused on time-integrated properties of networks~\cite{holme11-1,holme2015modern}. As result, complex temporal dynamics acting at shorter time-scales have been traditionally discarded. However, the recent technological advances in data storing and collection are providing unprecedented means to probe also the temporal dimension of real systems. The access to this feature is allowing to discover properties of social acts invisible in time aggregated datasets, and is helping characterizing the microscopic mechanisms driving their dynamics at all time-scales~\cite{perra12-1,karsai13-1,ubaldi2015asymptotic,laurent2015calls,moro10-1,clauset07,Isella:2011,saramaki2015seconds,Saramaki21012014,Sekara06092016}.  The advances in this arena are allowing to investigate the effects such temporal dynamics have on dynamical processes unfolding on time-varying networks. The study of the propagation of infectious diseases, idea, rumours, or memes etc.. on temporal graphs shows a rich and non trivial phenomenology radically different than what is observed on their static or annealed counter parts~\cite{barrat2015face,perra12-2,perra12-1,ribeiro12-2,PhysRevLett.112.118702,PhysRevE.87.032805,10.1063,starnini13-1,starnini_rw_temp_nets,valdano2015analytical,scholtes2014causality,Williams160196,rocha2014random,takaguchi2012importance,rocha2013bursts,ghoshal2006attractiveness,sun2015contrasting,mistry2015committed,pfitzner13-1,takaguchi12-1,takaguchi2013bursty,holme2014birth,holme2015basic,wang2016statistical,gonccalves2015social}.\\
{\indent}Before going any further, it is important to notice how in their more general form, multilayers networks, might be characterized by different types of nodes in each layers. For example, modern transportation systems in cities can be characterized as a multilayer network in which each layer captures a different transportation mode (tube, bus, public bikes etc..) and the links between layers connect stations (nodes) where people can switch mode of transport~\cite{boccaletti2014structure,de2016physics}. A particular version of multilayer networks, called multiplex, is typically used in social networks. Here, the entities in each layers are of the same type (i.e., people). The inter-layer links are drawn only to connect the same person in different layers.\\
{\indent}In this context, we introduce a model of time-varying multiplex networks. We aim to characterize the effects of temporal connectivity patterns and multiplexity on contagion processes. We model the intra-layer evolution of connections using the activity-driven framework~\cite{perra12-1}. In this model of time-varying networks, nodes are assigned with an activity describing their propensity to engage in social interactions per unit time~\cite{perra12-1}. Once active a node selects a partner to interact. Several selection mechanisms have been proposed, capturing different features of real social networks~\cite{karsai13-1,ubaldi2015asymptotic,ubaldi2016burstiness,alessandretti2017random,nadini2018epidemic}. The simplest, that will be used here, is memoryless and random~\cite{perra12-1}. The overlap between layers is modulated by a probability $p$. If $p=1$ all nodes are present in all layers. If $p=0$, the multiplex is formed by $M$ disconnected graphs. We consider $p$ as parameter and explore different regime of coupling between layers. Furthermore, each layer is characterized by an activity distribution. We consider different scenarios in which the activity of overlapping nodes (regulated by $p$) is uncorrelated and others in which is instead correlated. In these settings, we study the unfolding of Susceptible-Infected-Susceptible (SIS) epidemic processes~\cite{keeling08-1,alex12-1,pastor2015epidemic}. We derive analytically the epidemic threshold for two layers for any $p$ and any distributions of activities. In the limit of $p=1$ we find analytically the epidemic threshold for any number of layers. Interestingly, the threshold is a not trivial generalization of the correspondent quantity in the monoplex (single layer network). In the general case $0<p<1$ we found that the threshold is decreasing function of $p$. Positive correlations of overlapping nodes push the threshold to smaller values respect to the uncorrelated and negatively correlated cases. Furthermore, when the average connectivity of two layers is very different the critical behaviour of the system is driven by the more densely connected layer. In such scenario the epidemic threshold is not affected by the multiplexity, its value is equivalent to the case of a monoplex, and the overlap affects only the layer featuring the smaller average connectivity.\\
{\indent}The paper is organized as follow. In Section~\ref{model} we introduce the multiplex model. In Section~\ref{SIS_p1} we study first both analytically and numerically the spreading of SIS processes. Finally, in Section~\ref{conclu} we discuss our conclusions. 

\section{Time-varying multiplex network model}
\label{model}

We first introduce the multiplex model. For simplicity of discussion, we will consider the simplest case in which the system is characterized by $M=2$ layers $A$ and $B$. However, the same approach can be used to create a multiplex with any number of layers. Let us define $N$ as the total number of nodes in each layer. In general, we have three different categories of nodes: $N^A$, $N^B$ and $N^o$. They describe, respectively, the number of nodes that are present only in layer $A$, $B$, or in both. The last category is defined by a parameter $p$: the overlap between layers. Thus, on average, we have $N^A=N^B=(1-p)N$ and  $N^o=pN$. As mentioned in the introduction, the temporal dynamics in each layer are defined by the activity-driven framework~\cite{perra12-1}. Thus, each non-overlapping node is characterized by an activity extracted from a distribution $f_A(a)$ or $f_B(a)$ which captures its propensity to be engaged in a social interaction per unit time. Observations in real networks show that activity is typically heterogeneously distributed~\cite{perra12-1,karsai13-1,ubaldi2015asymptotic,ubaldi2016burstiness,tomasello2014role,ribeiro12-2}. Here, we assume that activities follow power-laws, thus $f_\bb{x}(a)=c_\bb{x} a^{-\gamma_\bb{x}}$ with $\bb{x}=[A,B]$ and $\epsilon \le a \le 1$ to avoid divergences. The overlapping nodes instead, are characterized by a joint activity distribution $h(a_A, a_B)$. As mentioned in the introduction, real multiplex networks are characterized by correlations across layers~\cite{starnini2017effects}. To account for such feature, we will consider three prototypical cases in which the activities of overlapping nodes in the two layers are i) uncorrelated ii) positively and iii) negatively correlated. To simplify the formulation and to avoid adding other parameters, in case of positive and negative correlations we adopt the following steps. We first extract the activities of the overlapping nodes from the two distributions $f_\bb{x}(a)$. Then we order them. In the case of positive correlation, a node that has the $r^{th}$ activity in $A$ will be assigned to the correspondent activity in $B$. In the case of negative correlations instead, a node that has the $r^{th}$ activity in $A$ will be assigned the $(pN-r+1)^{th}$ in $B$. \\

\begin{figure}
\includegraphics[width=0.4\textwidth]{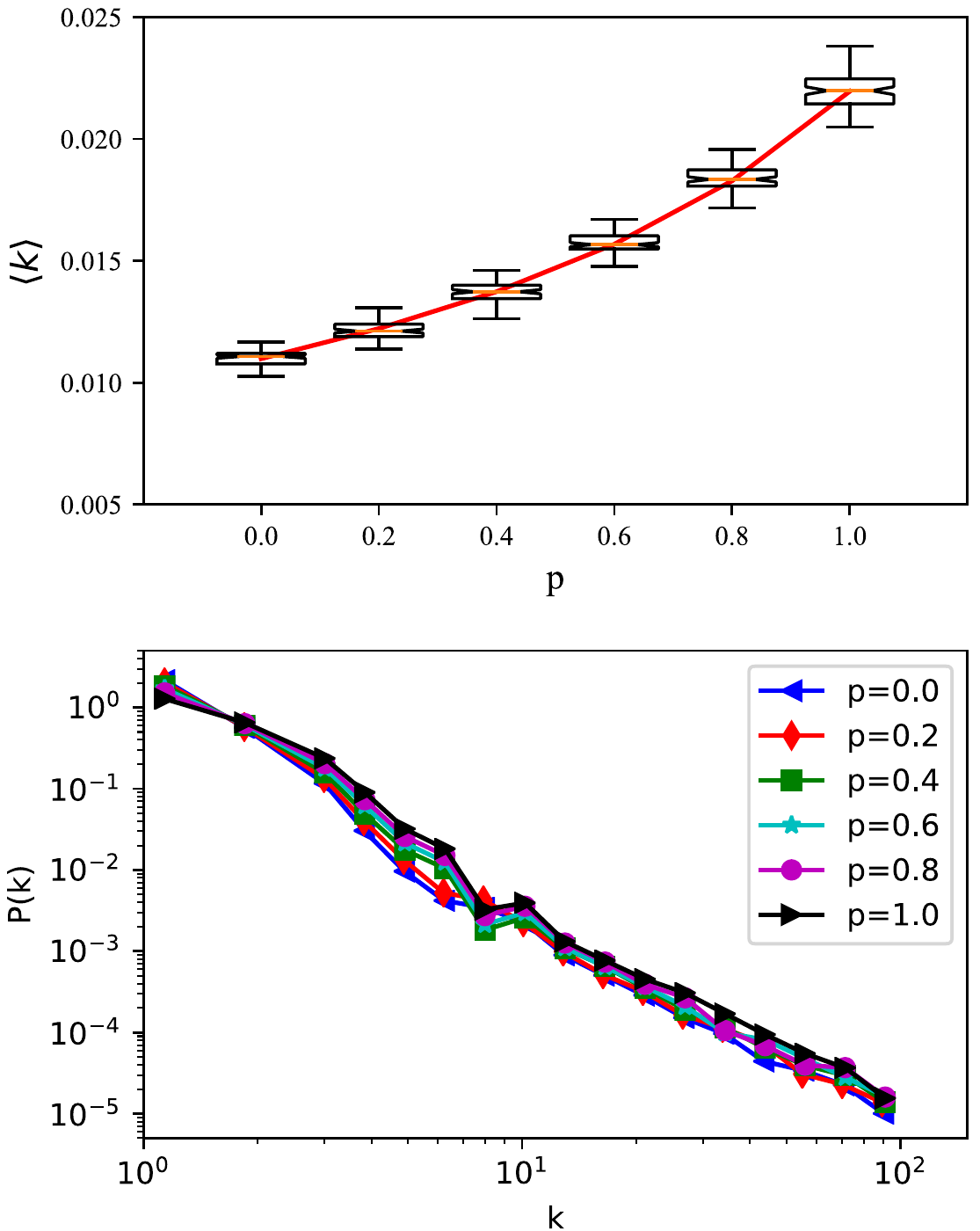}
\caption{(Top panel) The average overlapping degree of the time-integrated multiplex networks as function of the fraction of the overlapping nodes, $p$. The red line is computed by Eq.~(\ref{ave_k}). (Bottom panel) The distribution of the integrated overlapping degree for different $p$. Both in panels, the activities of the overlapping nodes in the two layers are uncorrelated. The exponents for the distributions of activities are $\gamma_\bb{x}=2.1$. The simulations are performed on networks of $10^5$ nodes with $m_\bb{x}=1$, integrated over $100$ time steps and averaged over 100 runs.}
\label{figure:ave_p_k}
\end{figure}

In these settings, the temporal evolution of the multiplex is defined as follow. For each realization, we randomly select $pN$ nodes as overlapping nodes between two layers. At each time step $t$:
\begin{itemize}
\item Each node is active with a probability defined by its activity. 
\item Each active node creates $m_\bb{x}$ links with randomly selected nodes. Multiple connections with the same node in the same layer within each time step are not allowed.
\item Overlapping nodes can be active and create connections in both layers.
\item At time step $t+\Delta t$ all connections are deleted and the process restart from the first point.
\end{itemize}
Connections have all the same duration of $\Delta t$. In the following, we set, without lack of generality, $\Delta t=1$. At each time the topology within each layer is characterized, mostly, by a set of disconnected stars of size $m_\bb{x}$. Thus, at the minimal temporal resolution each network looks very different than the static or annealed graphs we are used to see in the literature~\cite{newman10-1}. However, it is possible to show that, integrating links over $T$ time steps in the limit in which $T\ll N$, the resulting network has a degree distribution that follows the activity~\cite{perra12-1,starnini13-2,ubaldi2015asymptotic}. This is qualitatively similar to what is observed in real temporal networks where the topological features at different time-scales are very different than the late (or time integrated) characteristics~\cite{Sekara06092016}.\\
{\indent}At each time step the average degree in each layer can be computed as:
\be
\av{k}^\bb{x}_t=\frac{2E^\bb{x}_t}{N}=2m_\bb{x}\left [(1-p)\av{a_\bb{x}}+p\av{a_\bb{x}}_o \right ],
\ee
where $E^\bb{x}_t$ is the number of links generated in each layer at each time step. Furthermore, $\av{a_\bb{x}}=\sum_{a}f_\bb{x}(a)a$ and $\av{a_\bb{x}}_o=\sum_{a_A}\sum_{a_B}h(a_A,a_B)a_\bb{x}$ are the average activity of non-overlapping and overlapping nodes in each layer respectively. Similarly, the total average degree (often called overlapping degree~\cite{battiston2014structural}), at each time step, is:
\be
\label{ave_k}
\av{k}_t=\frac{2\sum \limits_{y \in \bb{x} }E^y_t}{2N -pN}=2\frac{\sum \limits_{y \in \bb{x} }m_y \left [ (1-p)\av{a_y}+p \av{a_y}_o \right ] }{(2-p)}.
\ee
Thus, the average connectivity, at each time step, is determined by the number of links created in each layer, and by the interplay between the average activity of overlapping and non-overlapping nodes. As shown in Figure~\ref{figure:ave_p_k} (top panel), Eq.~\ref{ave_k} describes quite well the behaviour of the average overlapping degree which is an increasing function of $p$. Indeed, the larger the fraction of overlapping nodes, the larger the connectivity of such nodes across layers. As we will see in the next section, this feature affects significantly the unfolding on contagion processes. \\
{\indent}In Figure~\ref{figure:ave_p_k} (bottom panel) we show the integrated degree distribution of the overlapping degree for different $p$. The plot clearly shows how the functional form is defined by the activity distributions of the two layers which in this case are equal. An increase in the fraction of overlapping nodes, does not change the distribution of the overlapping degree, it introduces a vertical shift which however is more visible for certain values of $k$. \\

\section{Contagion processes}
\label{SIS_p1}

In order to understand how the interplay between multiplexity and temporal connectivity patterns affects dynamical processes, we consider SIS contagion phenomena spreading on the multiplex model introduced in the previous section. In this prototypical epidemic model each node can be in one of two compartments. Healthy nodes are susceptible to the disease and thus in the compartment $S$. Infectious nodes instead join the compartment $I$. The natural history of the disease is defined as follows. A susceptible, in contact with an infected node, might get sick and infectious with probability $\lambda$. Each infected node spontaneously recovers with rate $\mu$ thus staying infectious for $\mu^{-1}$ time steps, on average. One crucial feature of epidemic models is the threshold which determines the conditions above which a disease is able to affect a macroscopic fraction of the population~\cite{keeling08-1,alex12-1,pastor2015epidemic}. In case of SIS models, below the threshold the disease dies out reaching the so called disease-free equilibrium. Above threshold instead, the epidemic reaches an endemic stationary state. This can be captured running the simulations for longer times and thus estimating the fraction of infected nodes for $t\rightarrow \infty$: $i_\infty$. In general, in a multiplex network, such fraction might be different across layers. Thus, we can define: $i_\infty^\bb{x}$. To characterize the threshold we could study the behavior of such fraction(s) as function of $\lambda/\mu$. Indeed, the final number of infected nodes acts as order parameter in a second order phase transition~\cite{alex12-1}. However, due to the stochastic nature of the process, the numerical estimation of the endemic state, especially in proximity of the threshold is not easy. Thus, we adopt another method measuring the life time of the process, $L$~\cite{boguna13-1}. This quantity is defined as the average time the disease needs either to die out or to infect a macroscopic fraction $Y$ of the population. The life time acts as the susceptibility in phase transition thus allows a more precise numerical estimation~\cite{boguna13-1}.\\
{\indent}In the case of single layer activity-driven networks, in which partners of interactions are chosen at random and without memory of past connections, the threshold can be written as (see Ref.~\cite{perra12-1} for details):
\be
\label{single_layer}
\frac{\lambda}{\mu}> \frac{1}{m}\frac{1}{\av{a}+\sqrt{\av{a^2}}}.
\ee
Thus, the conditions necessary for the spread of the disease are set by the interplay between the features of the disease (left side) and the dynamical properties of the time-varying networks where the contagion unfolds (right side). The latter are regulated by first and second moment of the activity distribution and by the number of connections created by each active node (i.e., $m$). It is important to notice that Eq.~\ref{single_layer} considers the case in which the time-scale describing the evolution of the connectivity patterns and the epidemic process are comparable. The contagion process is unfolding on a time-varying network. In the case when links are integrated over time and the SIS process spreads on a static or annealed version of the graph, the epidemic threshold will be much smaller~\cite{morris93-1,morris95-1,perra12-1}. This is due to the concurrency of connections which favours the spreading. In this limit of time-scale separation between the dynamics \emph{of} and \emph{on} networks, the evolution of the connectivity patterns is considered either much slower (static case) or much faster (annealed case) respect to the epidemic process. In the following, we will only consider the case of comparable time-scales.\\
{\indent}What is the threshold in the case of our multiplex and time-varying network model? In the limit $p=0$ the number of overlapping nodes is zero. The two layers are disconnected thus the system is characterized by two independent thresholds regulated  by the activity distributions of the two layers. The most interesting question, is then what happens for $p>0$. To find an answer to this conundrum, let us define $I^\bb{x}_a$ ($\bb{x}=[A,B]$) as the number of infected nodes of activity class $a$ that are present only in layer A or B. Clearly, $I^\bb{x}=\sum_{a}I^\bb{x}_a$. Let us instead define $I^o_{a_A, a_B}$ the number of infected overlapping nodes in classes of activity $a_A$ and $a_B$. In this case the total number of infected overlapping nodes is $I^o=\sum_{a_A}\sum_{a_B}I^o_{a_A, a_B}$. Similarly, we can define $N^\bb{x}_a$ and $N^o_{a_A,a_B}$ as the number of nodes non-overlapping nodes of activity $a$ and as the number of overlapping nodes of activity $a_A$ and $a_B$. The implicit assumption we are making by dividing nodes according to their activities, is that of statistical equivalence within activity classes~\cite{barrat08-1,alex12-1}. In these settings, we can write the variation of the number of infected non-overlapping nodes as function of time as: 
\bea \label{IAX}
&&d_t I_a^\bb{x} = -\mu I_a^\bb{x}+\lambda m_\bb{x} \left [ N_a^\bb{x}-I_a^\bb{x} \right ]a \frac{I^\bb{x}+I^o}{N}\\ \nonumber
&&+\lambda m_\bb{x} \frac{N_a^\bb{x}-I_a^\bb{x}}{N}\left [ \sum \limits_{a'}I_{a'}^\bb{x} {a'}+ \sum \limits_{{a'}_A}\sum \limits_{{a'}_B}I_{{a'}_A,{a'}_B}^o {a'}_\bb{x}\right ],
\eea
where we omitted the dependence of time. The first term on the r.h.s. considers nodes recovering thus leaving the infectious compartment. The second and third terms account for the activation of susceptibles in activity class $a$ ($S_a^\bb{x}=N_a^\bb{x}-I_a^\bb{x}$) that select as partners infected nodes (non-overlapping and overlapping) and get infected. The last two terms instead consider the opposite: infected nodes activate, select as partners non-overlapping and overlapping nodes in the activity class $a$ infecting them as result. \\
{\indent}Similarly, we can write the expression for the variation of overlapping nodes of activity classes $a_A$ and $a_B$ as:
\bea \label{IABO}
&&d_t I_{a_A,a_B}^o = -\mu I_{a_A,a_B}^o \\ \nonumber
&&+ \lambda \left [ N_{a_A,a_B}^o-I_{a_A,a_B}^o \right ]\sum \limits_{y \in \bb{x} } m_y a_y \frac{I^y+I^o}{N} \\ \nonumber
&& +\lambda \frac{N_{a_A,a_B}^o-I_{a_A,a_B}^o}{N} \\ \nonumber
&&\times\sum \limits_{y \in \bb{x} } m_y \left [ \sum \limits_{{a'}}I_{a'}^y {a'}+ \sum \limits_{{a'}_A}\sum \limits_{{a'}_B}I_{{a'}_A,{a'}_B}^o {a'}_y\right ].
\eea
The general structure of the equation is similar to the one we wrote above. The main difference is however that overlapping nodes can be infected and can infect in both layers. The first term in the r.h.s. accounts for the recovery process. The next four (two for each element in the sum in $y$) consider the activation of susceptible nodes that select as partners both non-overlapping and overlapping infected nodes and get infected. The last four terms account for the reverse process. In order to compute the epidemic threshold we need to define four auxiliary functions thus defining a closed system of differential equations. In particular, we define $\Theta^\bb{x}=\sum_{a}I^\bb{x}_a a$ and $\Theta^o_\bb{x}=\sum_{a_A}\sum_{a_B}I^o_{a_A,a_B} a_\bb{x}$. For simplicity, we will skip the detailed derivation here (see the Appendix for the details). By manipulating the previous three differential equations we can obtain four more, one for each auxiliary function. The condition for the spreading of the disease can be obtained by study the spectral properties of the Jacobian matrix of such system of seven differential equations.

\subsection{Two layers and $p=0$}
As sanity check, let us consider first the limit $p=0$. In this case, each layer acts independently and we expect the threshold of each to follow Eq.~\ref{single_layer}. This is exactly what we find. In particular, two of the seven eigenvalues are
\be
\Lambda_\bb{x}= -\mu +\lambda m_\bb{x}\av{a_\bb{x}} +\lambda m_\bb{x} \sqrt{\av{a_\bb{x}^2}},
\ee     
where $\av{a_\bb{x}^n}=\sum_{a}f_\bb{x}(a)a^n$. Thus, the spreading process will be able to affect a finite fraction of the total population in case either of these two eigenvalues is larger than zero, which implies $\frac{\lambda}{\mu}> (m_\bb{x}\av{a_\bb{x}}+m_\bb{x}\sqrt{\av{a_\bb{x}^2}})^{-1}$ as expected. It is important to notice that in case of a multiplex network the disease might be able to spread in one layer but not in the other. However, in case the condition for the spreading is respected in both layers, they will experience the disease.

\subsection{Two layers and $p=1$}

\begin{figure}
\includegraphics[width=0.5\textwidth]{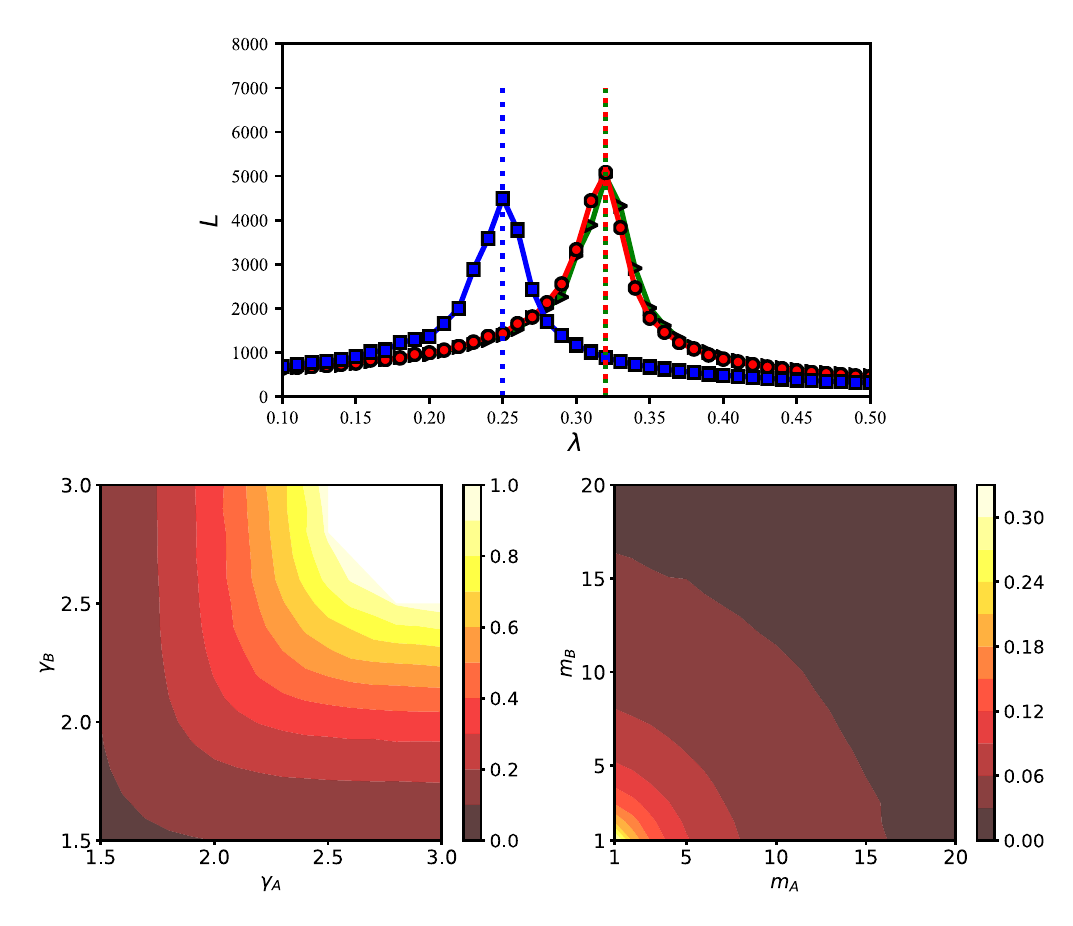}
\caption{(Top panel) Lifetime of SIS processes on temporal multiplex networks versus $\lambda$ for the uncorrelated (red circles), negative correlated (green triangles) and positively correlated (blue squares) cases. Therein, $\gamma_\bb{x}=2.1$ and $m_\bb{x}=1$. The vertical dotted lines with the same colors to the simulations results are the corresponding analytical values obtained from Eq.~(\ref{two_layer}). The simulation results are averaged by $10^2$ runs. (Bottom left panel) The analytical thresholds computed by Eq.~(\ref{two_layer}) in the 2-D plane ($\gamma_A$,$\gamma_B$) when $m_\bb{x}=1$. (Bottom right panel) The analytical thresholds are calculated from Eq.~(\ref{two_layer}) in the 2-D plane ($m_A$,$m_B$) when $\gamma_\bb{x}=1$. In both bottom panels we considered uncorrelated activities.}
\label{figure:p1}
\end{figure}

Let us consider the opposite limit: $p=1$. As described in details in the Appendix, the condition for the spreading of the disease reads:
\be
\label{two_layer}
\frac{\lambda}{\mu}> \frac{1}{\sum \limits_{y \in \bb{x} }m_y\av{a_y}_o+\sqrt{2 m_A m_B\av{a_Aa_B}_o+\sum \limits_{y \in \bb{x} }m_y^2\av{a_y^2}_o}},
\ee
where $\av{a_\bb{x}^n}_o=\sum_{a_A}\sum_{a_B}h(a_A,a_B)a_\bb{x}^n$ and $\av{a_A a_B}_o=\sum_{a_A}\sum_{a_B}h(a_A,a_B)a_A a_B$ . Interestingly, the threshold is function of the first and second moment of the activity distributions of the overlapping nodes which are modulated by the number of links each active node creates, plus a term which encode the correlation of the activities of such nodes in the two layers.\\
{\indent}Before showing the numerical simulations to validate the mathematical formulation an important observation is in order. In this limit, effectively, we could think the multiplex as a multigraph: a single layer network with two types of edges. In case the joint probability distribution of activity is $h(a_A,a_B)=f(a_A)\delta(a_B-a_A)$, thus two activities are exactly the same, and $m_A=m_B$ the threshold reduces to Eq.~\ref{single_layer} (valid for a single layer network) in which the number of links created by active nodes is $2m$. However, for a general form of the joint distribution and in case of different number of links created by each active node in different layers this correspondence breaks down.\\
{\indent}In all the following simulations, we set $N=10^5$, $\epsilon=10^{-3}$, $\mu=0.015$, $Y=0.3$, start the epidemic process from a $1\%$ of nodes selected randomly as initial seeds, and show the averages of $10^2$ independent simulations. In Figure~\ref{figure:p1} we show the first results considering a simple scenario in which $m_A=m_B=1$ and the exponents for the distributions of activities are the same $\gamma_\bb{x}=2.1$. 
The first observation is that in all three cases the analytical solutions (vertical dotted lines) agree with the results from simulations. The second observation is that in case of positive correlation between the activities of nodes in two layers, the threshold is significantly smaller than in the other two cases. This is not surprising as the nodes sustaining the spreading in both layers are the same. Thus, effectively, active nodes are capable to infect the double number of other nodes. The thresholds of the uncorrelated and negatively correlated cases are very similar. In fact, due to the heterogeneous nature of the activity distributions, except for few nodes in the tails, the effective difference between the activities matched in reverse or random order is not large, for the majority of nodes. In Figure~\ref{figure:p1} (bottom panels) we show the behavior of the threshold as function of the activity exponents and the number of links created by active nodes in the two layers. For a given distribution of activity in a layer, increasing the exponent in the other results in an increase of the threshold. This is due to the change of the first and second moments which decrease. In the settings considered here, if both exponents of activity distributions are larger than $2.6$ the critical value of $\lambda$ becomes larger than $1$, as shown in Figure~\ref{figure:p1} (left bottom). Thus, in such region of parameters, the disease will not spread. For a given number of links created in a layer by each active node, increasing the links created in the other layer results in a quite rapid reduction of the threshold. This is due to the increase of the connectivity and thus the spreading potential of active nodes. 

\subsection{$M$ layers and $p=1$}
\begin{figure}
\includegraphics[width=0.4\textwidth]{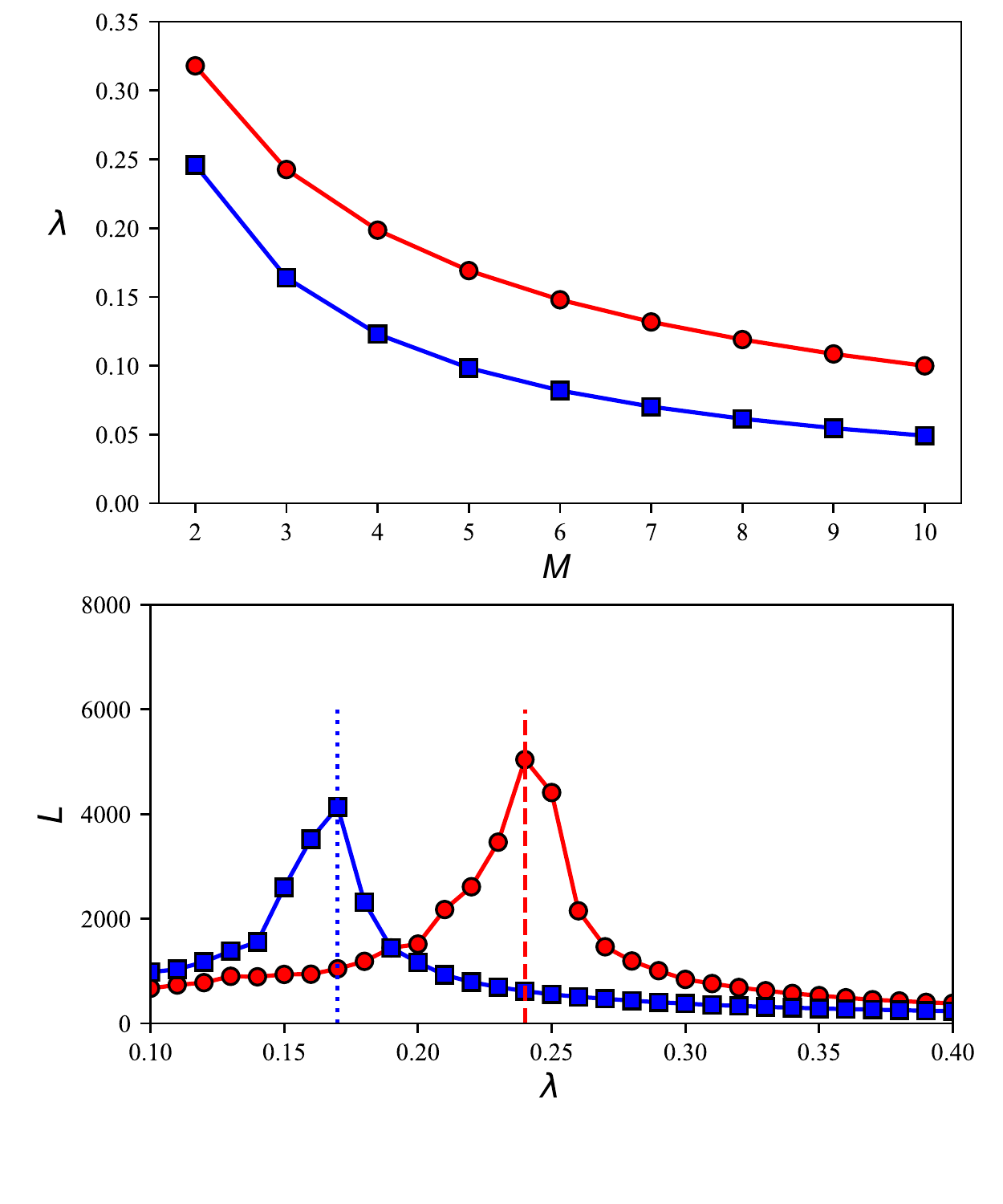}
\caption{The top panel shows the analytical threshold computed from Eq.~(\ref{M_layers}) versus the number of layers $M$ for the uncorrelated (red circles) and the positively correlated (blue squares) cases. The bottom panel presents the lifetime of SIS processes on temporal multiplex networks when $M=3$ for the uncorrelated (red circles) and the positively correlated (blue squares) cases. The blue dotted line and the red dashed line are the corresponding analytical values. Other parameters are set as $\gamma_\bb{x}=2.1$, $p=1$ and $m_\bb{x}=1$.}
\label{figure:p1_Ms}
\end{figure}

In the limit $p=1$, we are able to obtain an expression for the threshold of an SIS process unfolding on $M$ layers. The analytical condition for the spreading of the disease can be written as (see the Appendix for details):
\be
\footnotesize
\label{M_layers}
\frac{\lambda}{\mu}> \frac{1}{\sum \limits_{y\in \bb{x}}\av{a_y}_om_y+\sqrt{\sum \limits_{y\in \bb{x}}\av{a^2_y}_o m^2_y+\sum \limits_{y=A}^{M-1}\sum \limits_{z>y}^{M}2\av{a_ya_z}_om_ym_z}},
\ee
where $\bb{x}=[A,B,\ldots, Z]$ and $z>y$ implies an alphabetical ordering. The first observation is that in case $h(a_y,a_z)=f(a_y)\delta(a_z-a_y)\;\ \forall y,z \in \bb{x}$, thus the activity is the same for each node across each layer, Eq.~\ref{M_layers} reduces to:
\be
\frac{\lambda}{\mu}> \frac{1}{M m}\frac{1}{\av{a}+\sqrt{\av{a^2}}},
\ee
which is the threshold for a single layer activity-driven network in which $m \rightarrow Mm$. This is the generalization of the correspondence between the two thresholds we discussed above for two layers. The second observation is that, in general, increasing the number of layers decreases the epidemic threshold. Indeed, each new layers increases the connectivity potential of each node and thus the fragility of the system to the contagion process. Figure~\ref{figure:p1_Ms} (top panel) shows the analytical behavior of the epidemic threshold up to $M=10$ for the simplest case of uncorrelated (red dots) and positively correlated (blue squares) activities between layers confirming this result. In Figure~\ref{figure:p1_Ms} (bottom panel) we show the comparison between the analytical results and the numerical simulations. The plot shows a perfect match between the two. 

\subsection{Two layers and $0<p<1$}

We now turn the attention to the most interesting cases which are different from the two limits of null and total overlap of nodes considered above. For a general value of $p$, we could not find a general closed expression for the epidemic threshold. However, the condition for the spreading can be obtained by investigating, numerically, the spectral properties of the Jacobian (see the Appendix for details). In Figure~\ref{figure:ps} we show the lifetime of SIS spreading processes unfolding on a multiplex network for three different values of $p$. The top panel shows the uncorrelated case and the dashed vertical lines describe are the analytical predictions. The first observation is that the larger the overlap of nodes between two layers the smaller the threshold. Indeed, by increasing the fraction of nodes active in both layers increases the spreading power of such nodes when they get infected. The second observation is that the analytical predictions match remarkably well the simulations. The bottom panel shows instead the case of positive correlation between the activities of overlapping nodes in the two layers. Also in this case, the larger the overlap the smaller the epidemic threshold. However, the comparison between the two panels clearly show the effects of positive correlations. Indeed, for all the values of $p$ positive correlations push the threshold to smaller values respect to the uncorrelated case. This effect is larger for larger values of overlap. It is important to notice that also here our analytical predictions match remarkably well the numerical simulations. 
\begin{figure}
\includegraphics[width=0.4\textwidth]{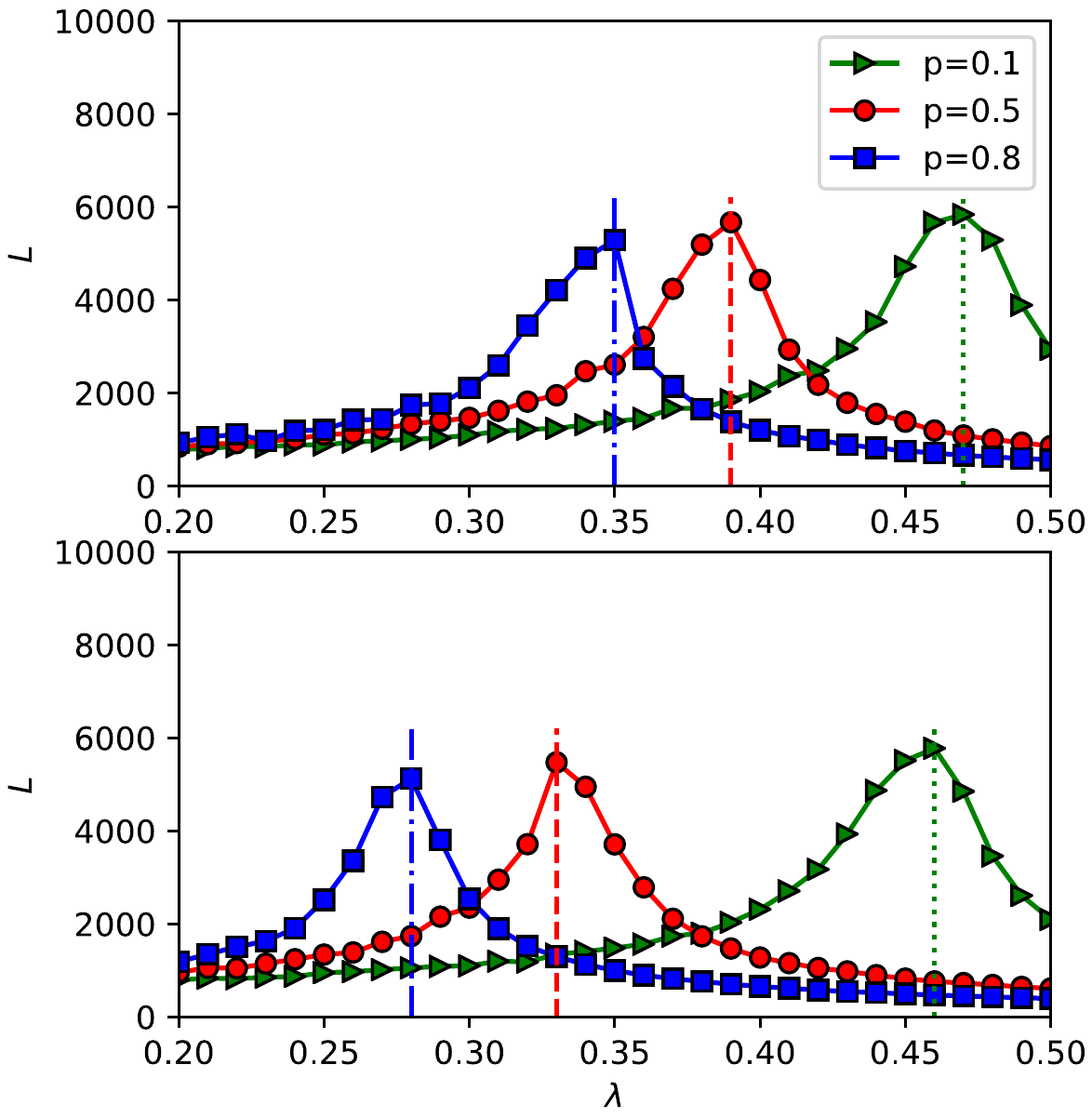}
\caption{Top panel shows the lifetimes of the SIS processes on the temporal multiplex networks in which the activities of the overlapping nodes in the two layers are uncorrelated for different fraction of overlapping nodes. Bottom panel shows the lifetimes of the SIS processes on the temporal multiplex networks in which the activities of the overlapping nodes in the two layers are positively correlated for different fraction of overlapping nodes. The vertical lines are the corresponding analytical values. Other parameters are set as $\gamma_\bb{x}=2.1$, $m_\bb{x}=1$.}
\label{figure:ps}
\end{figure}

\begin{figure}
\includegraphics[width=0.4\textwidth]{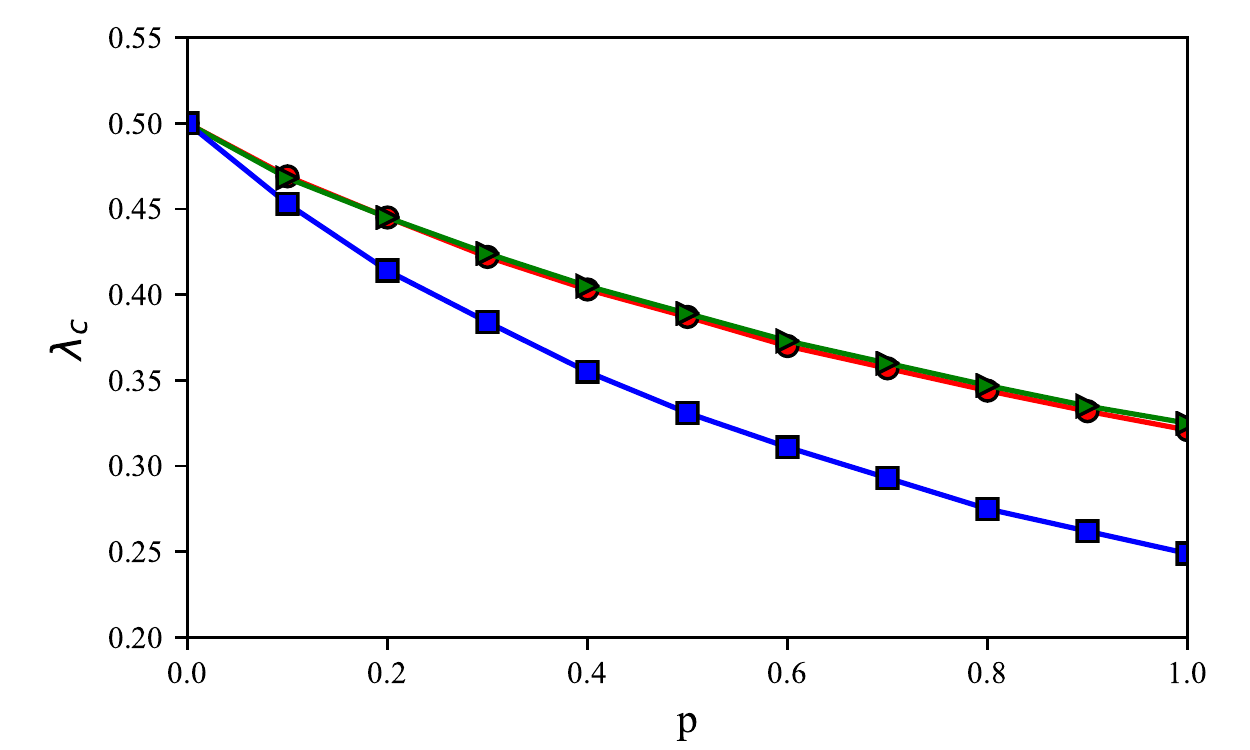}
\caption{The analytical threshold (Eq.~(\ref{J})) is plotted as a function of $p$ for the uncorrelated (red circles), negatively correlated (green triangles) and positively correlated (blue squares) cases. We set $\gamma_\bb{x}=2.1$ and $m_\bb{x}=1$. }
\label{figure:f(p)}
\end{figure}

In Figure~\ref{figure:f(p)} we show the behavior of the (analytical) epidemic threshold as function of $p$ for three types of correlations. The results confirm what discussed above. The larger the overlap the smaller the threshold. Negative and null correlations of overlapping nodes exhibit very similar thresholds. Instead, positive correlations push the critical value to smaller values. Furthermore, the larger the overlap the larger the effect of positive correlations as the difference between the thresholds increases as function of $p$. It is also important to notice how the threshold of a multiplex networks ($p>0$) is always smaller than the threshold of a monoplex ($p=0$) with the same features. Indeed, the presence of overlapping nodes effectively increases the spreading potential of the disease thus reducing the threshold. However, the presence of few overlapping nodes ($p \sim 0$) does not significantly change the threshold, this result and the effect of multiplexity on the spreading power of diseases is in line with what already discussed in the literature for static multiplexes~\cite{ginestra18-1,buono2014epidemics}.

\begin{figure}
\includegraphics[width=0.5\textwidth]{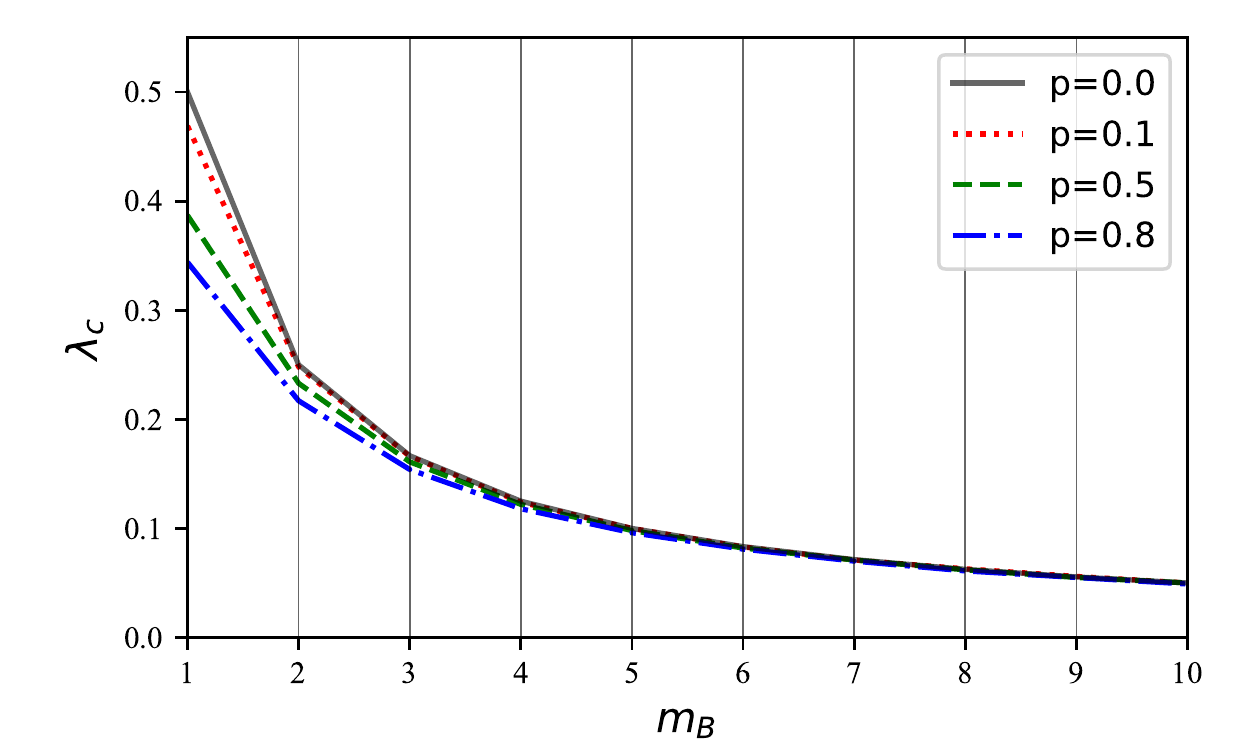}
\caption{The analytical threshold is plotted as a function of $m_B$ when the activities of the overlapping nodes in the two layers are uncorrelated and $m_A$ is fixed as 1. We set $\gamma_\bb{x}=2.1$.}
\label{figure:diffmb}
\end{figure}

In Figure~\ref{figure:diffmb}, we show how the epidemic threshold varies when the average connectivity of the two layers is progressively different and asymmetric. In other words, we investigate what happens when one layer has a much larger average connectivity than the other. This situation simulates individuals engaged in two different social contexts, one characterized by fewer interactions (e.g. close family interactions) and one instead by many more connections (e.g. work environment). In the figure, we consider a multiplex network in which the layer $A$ is characterized by $m_A=1$. We then let $m_B$ vary from $1$ to $10$ and measure the impact of this variation on the epidemic threshold for different values of $p$. For simplicity, we considered the case of uncorrelated activities in the two layers, but the results qualitatively hold also for the other types of correlations. Few observations are in order. As expected, the case $p=0$ is the upper bound of the epidemic threshold. However, the larger the asymmetry between the two layers, thus the larger the average connectivity in the layer $B$, the smaller the effect of the overlap on the threshold. Indeed, while systems characterized by $m_B=1$ and higher overlap feature a significantly smaller threshold respect to the monoplex, for $m_B \ge 3$ such differences become progressively negligible and the effects of multiplexity vanish. In this regime, the layer with the largest average connectivity drives the spreading of the disease. The connectivity of layer $B$, effectively determines the dynamics of the contagion, and thus the critical behavior is not influenced by overlapping nodes.

\begin{figure}
\includegraphics[width=0.4\textwidth]{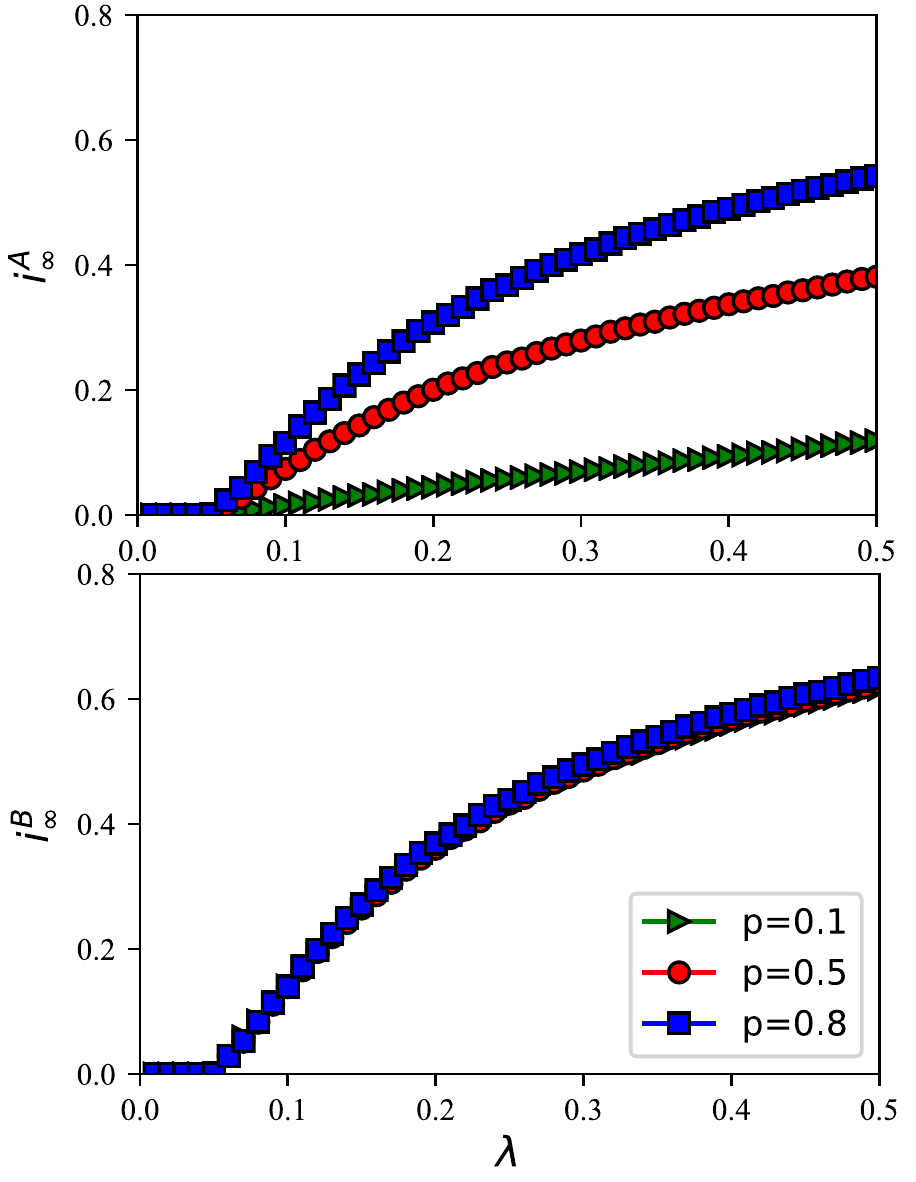}
\caption{Top panel and bottom panel respectively present the asymptotic fraction of infected nodes in layer $A$ and $B$. The results are averaged by $10^2$ simulations. We set $\gamma_\bb{x}=2.1$ and $m_\bb{x}=1$. }
\label{figure:inf_AB}
\end{figure}

In order to get a deeper understanding on this phenomena, we show the asymptotic number of infected nodes in each layers for $m_A=1$ and $m_B=10$ in Figure~\ref{figure:inf_AB}. For any $\lambda$ above the threshold the fraction of infected nodes in layer $B$ (bottom panel) is larger than in layer $A$ (top panel ) and is independent on the fraction of overlapping nodes. As discussed above, in these settings the layer $B$ is driving the contagion process and the imbalance between the connectivity patterns is large enough to behave as a monoplex. However, for layer $A$ the contagion process is still highly influenced by $p$. Indeed, as the fraction of overlapping nodes increases layer $A$ is more and more influenced by the contagion process unfolding in $B$. Overall, these results are qualitatively similar with the literature of spreading phenomena in static multilayer networks~\cite{gomez2013diffusion}.

\section{Conclusions}
\label{conclu}

We presented a time-varying model of multiplex networks. The intra-layer temporal dynamics follow the activity-driven framework which was developed for single layered networks (i.e. monoplexes). Thus, nodes are endowed with an activity that describes their propensity, per unit time, to initiate social interactions.  A fraction $p$ of nodes is considered to be overlapping between layers and their activities are considered, in general, to be different but potentially correlated. In these settings, we studied how multiplexity and temporal connectivity patterns affect dynamical processes unfolding on such systems. To this end, we considered a prototypical model of infectious diseases: the SIS model. We derived analytically the epidemic threshold of the process as function of $p$. In the limit $p=0$ the system is constituted by disconnected networks that behave as monoplexes. In the opposite limit instead ( i.e. $p=1$) the epidemic threshold is function of the first and second moment of the activity distributions as well as by their correlations across layers.  We found that, systems characterized by positive correlations are much more fragile to the spreading of the contagion process with respect to negative and null correlations. The threshold also varies as a function of the number of layers $M$. Indeed, with perfect overlapping, each node is present and active in each layer. Thus, the larger $M$ the smaller the epidemic threshold as the spreading potential of each node increases. In the general case $0<p<1$, we could not find a closed expression for the epidemic threshold. However, the critical condition for the spreading can be calculated from the theory by investigating numerically the spectral properties of the Jacobian matrix describing the contagion dynamics. Also in this case, positive correlations of activities across layers help the spreading by lowering the epidemic threshold; while negative and null correlations result in very similar thresholds.  Moreover, the larger the overlap between layer the lower the critical condition for the spreading. Indeed, the case of disconnected monoplexes (i.e. $p=0$) is the upper bound for the threshold. Interestingly, the role of the overlap, thus of the multiplexity, is drastically reduced in case the average connectivity in one layer is much larger than the other. In this scenario, which mimics the possible asymmetry in the contact patterns typical of different social contexts (e.g. family VS work environment), one layer drives the contagion dynamics and the critical condition for the spreading is indistinguishable from the monoplex. However, the overlap is still significantly important in the other layer as the fraction of nodes present in both layers largely determines the spreading dynamics. 

Some of these results are qualitatively in line with the literature of contagion processes unfolding on static/annealed multiplexes. However, as known in the case of single layered graphs, time-varying dynamics induce large quantitative differences~\cite{holme11-1,holme2015modern}. Indeed, the concurrency and order of connections are crucial ingredients for the spreading and neglecting them, in favor of static/annealed representations, generally results in smaller thresholds. While the limits of time-scale separation might be relevant to describe certain types of processes, they might lead to large overestimation of the spreading potential of contagion phenomena. \\

The model presented here comes with several limitations. In fact, we considered the simplest version of the activity driven framework in which, at each time step, links are created randomly. Future work could explore the role of more realistic connectivity patterns in which nodes activate more likely a subset of (strong) ties and/or nodes are part of communities of tightly linked individuals. Furthermore, we assumed that the activation process is Poissonian and the activity of each node is not a function of time. Future work could explore most realistic dynamics considering bursty activation and ageing processes. All these features of real time-varying networks have been studied at length in the literature of single layered networks but their interplay with multiplexity when dynamical processes are concerned is still unexplored. Thus result presented here are a step towards the understanding of the temporal properties of multiplex networks and their impact on contagion processes unfolding on their fabric.

\section{ACKNOWLEDGMENTS}

Q.-H Liu would like to acknowledge M. Tang and W. Wang at the
Web Science Center for valuable discussions and comments at the beginning of this project.
Q.-H Liu acknowledges the support of the program of China Scholarships Council, National Natural Science Foundation of China (Grant No. 61673086) and Science Strength Promotion Programme of UESTC. We thank Alessandro Vespignani and Sandro Meloni for interesting discussions and comments. 

\section{Appendix}

\subsection{Derivation of the threshold when $M=2$}
Integrating over all activity spectrum of Eq.~(\ref{IAX}), it obtains the following equation,
\renewcommand{\theequation}{S\arabic{equation}}
\setcounter{equation}{0}
\bea \label{IX}
\nonumber
&&d_tI^\bb{x}=-{\mu}I^\bb{x}+\lambda m_\bb{x} \sum\limits_{a}\frac{N_a^\bb{x}-I_a^\bb{x}}{N}a(I^\bb{x}+I^o)\\ \nonumber
&&+\lambda m_\bb{x} \sum\limits_{a}\frac{N_a^\bb{x}-I_a^\bb{x}}{N}\bigg[ \sum\limits_{a^{\prime}}I_{a^{\prime}}^\bb{x} a^{\prime}+\sum\limits_{a_A^{\prime}}\sum\limits_{a_B^{\prime}}I_{a^{\prime}_A, a^{\prime}_B}^oa^{\prime}_\bb{x}\bigg].
\\
\eea
{\noindent}Initially, $I_a^\bb{x}\approx 0$, $\sum\limits_{a}\frac{N_a^\bb{x}-I_a^\bb{x}}{N}a\approx (1-p)\av{ a_\bb{x} }$. Thus Eq.~(\ref{IX}) can be further simplified as
\bea \label{IX_1}
\nonumber
&&d_tI^\bb{x} \approx-{\mu}I^\bb{x}+\lambda m_\bb{x} (1-p)\av{ a_\bb{x} }(I^\bb{x}+I^o)\\ \nonumber
&&+\lambda m_\bb{x} (1-p)\bigg[ \sum\limits_{a^{\prime}}I_{a^{\prime}}^\bb{x} a^{\prime}+\sum\limits_{a_A^{\prime}}\sum\limits_{a_B^{\prime}}I_{a^{\prime}_A, a^{\prime}_B}^oa^{\prime}_\bb{x}\bigg].
\\
\eea
{\noindent}Four auxiliary variables defined to simplify Eq.~(\ref{IX_1}) are as follows, $\Theta^A=\sum\limits_{a} I_{a}^{A}a$, $\Theta^B=\sum\limits_{a} I_{a}^{B}a$, $\Theta_{A}^{o}=\sum\limits_{a_B}\sum\limits_{a_A} I_{a_A, a_B}^{o}a_A$ and $\Theta_{B}^{o}=\sum\limits_{a_A}\sum\limits_{a_B} I_{a_A, a_B}^{o}a_B$. Since $\bb{x}=[A,B]$, Eq.~(\ref{IX_1}) can be expressed as
\bea \label{IA}
\nonumber
&&d_tI^A =-{\mu}I^A+\lambda m_A (1-p)\av{ a_A }(I^A+I^o)\\
&&+\lambda m_A (1-p)\big( \Theta^A+\Theta_A^o \big)
\eea
{\noindent}and 
\bea \label{IB}
\nonumber
&&d_tI^B =-{\mu}I^B+\lambda m_B (1-p)\av{ a_B }(I^B+I^o)\\
&&+\lambda m_B (1-p)\big(  \Theta^B+\Theta_B^o\big).
\eea
{\noindent}Integrating over all activity spectrum of Eq.~(\ref{IABO}), it obtains the following equation,
\bea \label{IO}
\nonumber
&&d_tI^o =-{\mu}I^o+\lambda m_A p \av{ a_A }_o (I^A+I^o)\\ \nonumber
&&+\lambda m_B p \av{ a_B }_o\big(  I^B+I^o\big)+p\lambda m_A (\Theta^A+\Theta_A^o)\\ 
&&+p\lambda m_B (\Theta^B+\Theta_B^o).
\eea
{\noindent}Multiplying both side of Eq.~(\ref{IAX}) by $a_\bb{x}$, and integrating over all activity spectrum, we get the following equation
\bea \label{thetaX}
\nonumber
&&d_t \Theta^\bb{x}=-\mu \Theta^\bb{x}+\lambda m_\bb{x} (1-p) \av{ a_\bb{x}^2 } \big (I^A+I^o \big) \\
&&+\lambda m_\bb{x} (1-p) \av{ a_\bb{x} } \big ( \Theta^\bb{x}+\Theta_\bb{x}^{o} \big ).
\eea
{\noindent}Replacing $\bb{x}$ with A and B in Eq.~(\ref{thetaX}) respectively, we have
\bea \label{thetaA}
\nonumber
&&d_t \Theta^A=-\mu \Theta^A+\lambda m_A (1-p) \av{ a_A^2 } \big (I^A+I^o \big)\\
&&+\lambda m_A (1-p) \av{ a_A } \big ( \Theta^A+\Theta_A^{o} \big ).
\end{eqnarray}
{\noindent}and
\bea \label{thetaB}
\nonumber
&&d_t \Theta^B=-\mu \Theta^B+\lambda m_B (1-p) \av{ a_B^2 } \big (I^B+I^o \big) \\
&&+\lambda m_B (1-p) \av{ a_B}\big ( \Theta^B+\Theta_B^{o} \big ).
\end{eqnarray}
{\noindent}In the same way, multiplying both sides of Eq.~(\ref{IABO}) by $a_\bb{x}$ and integrating over all activity spectrum, it obtains the following two equations
\bea \label{thetaAO}
\nonumber
&&d_t \Theta_A^{o}=-\mu \Theta_A^{o}+\lambda m_A p \av{ a_A^2}_o \big(I^A+I^o \big ) \\ \nonumber
&&+\lambda m_B p \av{a_A a_B}_o\big(I^B+I^o \big)+ \lambda m_A p \av{ a_A }_o \big(\Theta^A+\Theta_A^{o} \big)\\
&&+\lambda m_B p \av{a_A}_o \big(\Theta^B+\Theta_B^{o} \big)
\eea
\\
{\noindent}and
\bea \label{thetaBO}
\nonumber
&&d_t \Theta_B^{o} =-\mu \Theta_B^{o}+\lambda m_A p \av{ a_Aa_B }_o \big(I^A(t)+I^o(t) \big)\\ \nonumber
&&+\lambda m_B p \av{ a_B^2 }_o\big(I^B(t)+I^o(t) \big)+\lambda m_A p \av{ a_B }_o \big(\Theta^A+\Theta_A^{o} \big ) \\
&&+ \lambda m_B p \av{ a_B }_o \big(\Theta^B+\Theta_B^{o} \big)
\eea
{\noindent}when $\bb{x}$ is replaced with A and B, respectively. When the system enters the steady state, we have $d_t I^A=0$, $d_t I^B=0$, $d_t I^o=0$, $d_t \Theta^A=0$, $d_t \Theta^B=0$, $d_t \Theta_A^o=0$ and $d_t \Theta_B^o=0$. Set the right hand of Eqs. (\ref{thetaA})-(\ref{thetaBO}) and Eqs. (\ref{IA})-(\ref{IO}) as zero, and denote them respectively as F($\Theta^A$), F($\Theta^B$), F($\Theta_A^o$), F($\Theta_B^o$), F($I^A$), F($I^B$) and F($I^o$). Thus, the critical condition is determined by the following Jacobian matrix,
\begin{widetext}
\bea \label{J}
{J= 
\left[ \begin{array}{ccccccc}
\frac{\partial F(\Theta^A)}{\partial \Theta^A} & \frac{\partial F(\Theta^A)}{\partial \Theta^B} & \frac{\partial F(\Theta^A)}{\partial \Theta_A^o} &\frac{\partial F(\Theta^A)}{\partial \Theta_B^o} & \frac{\partial F(\Theta^A)}{\partial I^A} & \frac{\partial F(\Theta^A)}{\partial I^B} & \frac{\partial F(\Theta^A)}{\partial I^o} \\
\frac{\partial F(\Theta^B)}{\partial \Theta^A} & \frac{\partial F(\Theta^B)}{\partial \Theta^B} & \frac{\partial F(\Theta^B)}{\partial \Theta_A^o} &\frac{\partial F(\Theta^B)}{\partial \Theta_B^o} & \frac{\partial F(\Theta^B)}{\partial I^A} & \frac{\partial F(\Theta^B)}{\partial I^B} & \frac{\partial F(\Theta^B)}{\partial I^o} \\
\frac{\partial F(\Theta_A^o)}{\partial \Theta^A} & \frac{\partial F(\Theta_A^o)}{\partial \Theta^B} & \frac{\partial F(\Theta_A^o)}{\partial \Theta_A^o} &\frac{\partial F(\Theta_A^o)}{\partial \Theta_B^o} & \frac{\partial F(\Theta_A^o)}{\partial I^A} & \frac{\partial F(\Theta_A^o)}{\partial I^B} & \frac{\partial F(\Theta_A^o)}{\partial I^o} \\
\frac{\partial F(\Theta_B^o)}{\partial \Theta^A} & \frac{\partial F(\Theta_B^o)}{\partial \Theta^B} & \frac{\partial F(\Theta_B^o)}{\partial \Theta_A^o} &\frac{\partial F(\Theta_B^o)}{\partial \Theta_B^o} & \frac{\partial F(\Theta_B^o)}{\partial I^A} & \frac{\partial F(\Theta_B^o)}{\partial I^B} & \frac{\partial F(\Theta_B^o)}{\partial I^o} \\
\frac{\partial F(I^A)}{\partial \Theta^A} & \frac{\partial F(I^A)}{\partial \Theta^B} & \frac{\partial F(I^A)}{\partial \Theta_A^o} &\frac{\partial F(I^A)}{\partial \Theta_B^o} & \frac{\partial F(I^A)}{\partial I^A} & \frac{\partial F(I^A)}{\partial I^B} & \frac{\partial F(I^A)}{\partial I^o} \\
\frac{\partial F(I^B)}{\partial \Theta^A} & \frac{\partial F(I^B)}{\partial \Theta^B} & \frac{\partial F(I^B)}{\partial \Theta_A^o} &\frac{\partial F(I^B)}{\partial \Theta_B^o} & \frac{\partial F(I^B)}{\partial I^A} & \frac{\partial F(I^B)}{\partial I^B} & \frac{\partial F(I^B)}{\partial I^o} \\
\frac{\partial F(I^o)}{\partial \Theta^A} & \frac{\partial F(I^o)}{\partial \Theta^B} & \frac{\partial F(I^o)}{\partial \Theta_A^o} &\frac{\partial F(I^o)}{\partial \Theta_B^o} & \frac{\partial F(I^o)}{\partial I^A} & \frac{\partial F(I^o)}{\partial I^B} & \frac{\partial F(I^o)}{\partial I^o} 
\end{array} 
\right ]}
\eea
\end{widetext}
{\noindent}If the largest eigenvalue of J is larger than zero, the epidemic will outbreak.  Otherwise, the epidemic will die out. Specifically, if $p=0$, two layers are independent and we can get the following two of seven eigenvalues,
\bea
\nonumber
\Lambda_A&=&-\mu+a_A\lambda m_A + \lambda m_A \sqrt{\av{ a_A^2 }}
\eea
{\noindent} and
\bea
\nonumber
\Lambda_B&=&-\mu+a_B\lambda m_B + \lambda m_B \sqrt{\av{ a_B^2 }},
\eea
{\noindent}which determine the dynamics on layer A and layer B, respectively. If $p=1$, the largest eigenvalue is
\bea
\nonumber
&&\Lambda=-\mu+\lambda\sum \limits_{y \in \bb{x} }m_y\av{a_y}_o \\
&&+\lambda\sqrt{2 m_A m_B\av{a_Aa_B}_o+\sum \limits_{y \in \bb{x}}m_y^2\av{a_y^2}_o}.
\eea
{\noindent}Further, the critical transmission rate is written as
\bea \nonumber
\lambda_c= \frac{\mu}{\sum \limits_{y \in \bb{x} }m_y\av{a_y}_o+\sqrt{2 m_A m_B\av{a_Aa_B}_o+\sum \limits_{y \in \bb{x} }m_y^2\av{a_y^2}_o}}.
\\
\eea
For $0<p<1$, we could not find a general analytical expression for the eigenvalues of $J$. However, the critical transmission rate can be found by finding the value of $\lambda$ leading the largest eigenvalue of $J$ to zero. In other words, rather the solving explicitly the characteristic polynomial $|J-\Lambda I |=0$ and defining the condition for the spreading $\max{\Lambda}>0$ as done above, we can determined the critical value of $\lambda$ as the value corresponding to the largest eigenvalue to be zero~\cite{baxter2016unified,zhuang2017clustering}. 

\subsection{Derivation the threshold for $M$ layers when $p=1$}

Assume there are $M$ layers, and let $N_{a_A,a_B,...,a_M}^o$ and $I_{a_A,a_B,...,a_M}^o(t)$ respectively be the number of nodes and the number of infected nodes with activities $(a_A, a_B, ..., a_M)$ in layers $(A, B, ...,M)$. With the same derivation method of  Eq.~(\ref{IABO}), the evolution equation of $I_{a_A,a_B,...,a_M}^o$ can be written as
\bea  \label{IM}
\nonumber
&&d_tI_{a_A,...,a_M}^o = -\mu  I_{a_A,...,a_M}^{o}\\
\nonumber
&&+ \sum\limits_{i=A}^{M}\lambda m_i \big [ N_{a_A,...,a_M}^{o}-I_{a_A,...,a_M}^{o} \big ] a_i  \frac{I^o}{N}\\ 
\nonumber
&&+ \frac{N_{a_A,...,a_M}^{o}-I_{a_A,...,a_M}^{o}}{N} \sum\limits_{i=A}^{M}\lambda m_i \sum\limits_{a_A^{\prime}} \cdots \sum\limits_{a_M^{\prime}} I_{a_A^{\prime},...,a_M^{\prime}}^{o} a_i^{\prime},
\\
\eea
{\noindent}therein, $I^o=\sum\limits_{a_A^{\prime}} \cdots \sum\limits_{a_M^{\prime}} I_{a_A^{\prime},...,a_M^{\prime}}^{o}$. For the simplicity, let
\bea \label{thetaOA}
\Theta_{i}^{o}=\sum\limits_{a_A^{\prime}} \cdots \sum\limits_{a_M^{\prime}} I_{a_A^{\prime},...,a_M^{\prime}}^{o} a_i^{\prime}.
\eea
{\noindent}Multiplying both sides of Eq.(\ref{IM}) by $a_i$ and integrating over all activity spectrum, it obtains the following equation
\bea \label{ThetaIO}
\nonumber
&&d_t\Theta_i^{o} \approx -\mu  \Theta_i^{o}+\sum\limits_{a_A^{\prime}} \cdots \sum\limits_{a_M^{\prime}}a_i^{\prime} \sum\limits_{j=A}^{M}\lambda m_j \frac{N_{a_A^{\prime},...,a_M^{\prime}}^o}{N}a_j^{\prime}  I^o\\
\nonumber
&&+\sum\limits_{a_A^{\prime}} \cdots \sum\limits_{a_M^{\prime}}\frac{N_{a_A^{\prime},...,a_M^{\prime}}}{N}a_i^{\prime}\sum\limits_{j=A}^{M} \lambda m_j \Theta_j^o\\
&=&-\mu  \Theta_i^{o}+ \sum\limits_{j=A}^{M}\lambda m_j \av{ a_i a_j }  I^o+\av{ a_i } \sum\limits_{j=A}^{M}\lambda m_j \Theta_j^o.
\eea
{\noindent}Integrating over all activity spectrum of Eq.~(\ref{IM}), it obtains the following equation
\bea\label{IMS} \nonumber
&& d_tI^o= -\mu  I^{o}+ \sum\limits_{i=A}^{M}\lambda m_i {\av{ a_i }} I^o+\sum\limits_{i=A}^{M}\lambda m_i \Theta_i^{o}.
\\
\eea
{\noindent}When the system enters the steady state, $d_tI^o=0$ and $d_t\Theta_i^o=0$ for $i=A,B,...,M$. Set the right side of Eqs.~(\ref{ThetaIO}) and (\ref{IMS}) as zero, and denote them respectively as $F_i(\Theta_i^o)$ and $F(I^o)$. Thus the critical condition is determined by the following Jacobian matrix
\bea
{J_M= \nonumber
\left[ \begin{array}{cccc}
\frac{\partial F_A(\Theta_A^o)}{\partial \Theta_A^o} & \frac{\partial F_A(\Theta_A^o)}{\partial \Theta_B^o} & \cdots &  \frac{\partial F_A(\Theta_A^o)}{\partial I^o} \\
\frac{\partial F_B(\Theta_A^o)}{\partial \Theta_A^o} & \frac{\partial F_B(\Theta_A^o)}{\partial \Theta_B^o} & \cdots &  \frac{\partial F_B(\Theta_A^o)}{\partial I^o} \\
\cdots & \cdots & \cdots & \cdots \\
\frac{\partial F(I^o)}{\partial \Theta_A^o} & \frac{\partial F(I^o)}{\partial \Theta_B^o} & \cdots &  \frac{\partial F(I^o)}{\partial I^o}
\end{array}
\right ].}
\eea
{\noindent}Further, the maximum eigenvalue of matrix $J_M$ can be calculated as
\bea \nonumber
&&\Lambda= -\mu+\sum\limits_{i=A}^{M} \lambda \av{ a_i }  m_i \\ \nonumber
&&+\sqrt{\sum\limits_{i=A}^{M}\lambda^2\av{ a_i^2 } m_i^2 +\sum\limits_{i=A}^{M-1}\sum\limits_{j>i}^{M}2\lambda^2\av{ a_i a_j } m_im_j} 
\\
\eea
{\noindent}Thus, the critical transmission rate is
\bea \nonumber
\lambda_c= \frac{\mu}{\sum\limits_{i=A}^{M} \av{ a_i }  m_i+{\sqrt{\sum\limits_{i=A}^{M}\av{a_i^2} m_i^2 +\sum\limits_{i=A}^{M-1}\sum\limits_{j>i}^{M}2\av{ a_i a_j } m_im_j}}} \\
\eea
{\noindent}Further, if the activities of the same node in each layer are the same, the above equation can be simplified as follows, 
\bea
\lambda_c= \frac{1}{Mm}\frac{\mu}{ \av{a} +{\sqrt{\av{ a^2 }}}}
\eea

\end{document}